\documentstyle[12pt]{article}

\catcode`\@=11
\def\@citex[#1]#2{%
\if@filesw \immediate \write \@auxout {\string \citation {#2}}\fi
\@tempcntb\m@ne \let\@h@ld\relax \def\@citea{}%
\@cite{%
  \@for \@citeb:=#2\do {%
    \@ifundefined {b@\@citeb}%
      {\@h@ld\@citea\@tempcntb\m@ne{\bf ?}%
      \@warning {Citation `\@citeb ' on page \thepage \space undefined}}%
      {\@tempcnta\@tempcntb \advance\@tempcnta\@ne%
      \@tempcntb\number\csname b@\@citeb \endcsname \relax%
      \ifnum\@tempcnta=\@tempcntb 
	\ifx\@h@ld\relax%
	  \edef \@h@ld{\@citea\csname b@\@citeb\endcsname}%
	\else%
	  \edef\@h@ld{\ifmmode{-}\else--\fi\csname b@\@citeb\endcsname}%
	\fi%
      \else
	\@h@ld\@citea\csname b@\@citeb \endcsname%
	\let\@h@ld\relax%
      \fi}%
    \def\@citea{,\penalty\@highpenalty\,}%
  }\@h@ld
}{#1}}

\def\@citeb#1#2{{[#1]\if@tempswa , #2\fi}}
\def\@citeu#1#2{{$^{#1}$\if@tempswa , #2\fi }}
\def\@citep#1#2{{#1\if@tempswa , #2\fi}}

\def\bcites{         
	\catcode`\@=11
	\let\@cite=\@citeb
	\catcode`\@=12
}

\def\upcites{         
	\catcode`\@=11
	\let\@cite=\@citeu
	\catcode`\@=12
}

\def\plaincites{      
	\catcode`\@=11
	\let\@cite=\@citep
	\catcode`\@=12
}

\def\section{\@startsection {section}{1}{\z@}{3.ex plus 1ex minus
 .2ex}{2.ex plus .2ex}{\large\bf}}
\def\subsection{\@startsection{subsection}{2}{\z@}{2.75ex plus 1ex minus
 .2ex}{1.5ex plus .2ex}{\bf}}

\def\appendix{{\newpage\section*{Appendices}}\let\appendix\section%
	{\setcounter{section}{0}
	\gdef\thesection{\Alph{section}}}\section}

\def\abstract{\if@twocolumn
\section*{Abstract}
\else 
\begin{center}
{\bf Abstract\vspace{-.5em}\vspace{0pt}}
\end{center}
\quotation
\fi}

\catcode`\@=12

\def\noj#1,#2,{{\bf #1} (19#2)\ }
\def\jou#1,#2,#3,{{\sl #1\/ }{\bf #2} (19#3)\ }
\def\ann#1,#2,{{\sl Ann.\ Physics\/ }{\bf #1} (19#2)\ }
\def\cmp#1,#2,{{\sl Comm.\ Math.\ Phys.\/ }{\bf #1} (19#2)\ }
\def\cq#1,#2,{{\sl Class.\ Quantum Grav.\/ }{\bf #1} (19#2)\ }
\def\cqg#1,#2,{{\sl Class.\ Quantum Grav.\/ }{\bf #1} (19#2)\ }
\def\ijmp#1,#2,{{\sl Int.\ J.\ Mod.\ Phys.\/ }{\bf A#1} (19#2)\ }
\def\jmp#1,#2,{{\sl J.\ Math.\ Phys.\/ }{\bf #1} (19#2)\ }
\def\lmp#1,#2,{{\sl Lett.\ Math.\ Phys.\/ }{\bf #1} (19#2)\ }
\def\grg#1,#2,{{\sl Gen.\ Rel.\ Grav.\/ }{\bf #1} (19#2)\ }
\def\mpl#1,#2,{{\sl Mod.\ Phys.\ Lett.\/ }{\bf A#1} (19#2)\ }
\def\nc#1,#2,{{\sl Nuovo Cim.\/ }{\bf #1} (19#2)\ }
\def\np#1,#2,{{\sl Nucl.\ Phys.\/ }{\bf B#1} (19#2)\ }
\def\pl#1,#2,{{\sl Phys.\ Lett.\/ }{\bf #1B} (19#2)\ }
\def\pla#1,#2,{{\sl Phys.\ Lett.\/ }{\bf #1A} (19#2)\ }
\def\pr#1,#2,{{\sl Phys.\ Rev.\/ }{\bf #1} (19#2)\ }
\def\prd#1,#2,{{\sl Phys.\ Rev.\/ }{\bf D#1} (19#2)\ }
\def\prl#1,#2,{{\sl Phys.\ Rev.\ Lett.\/ }{\bf #1} (19#2)\ }
\def\prp#1,#2,{{\sl Phys.\ Rept.\/ }{\bf #1C} (19#2)\ }
\def\ptp#1,#2,{{\sl Prog.\ Theor.\ Phys.\/ }{\bf #1} (19#2)\ }
\def\ptpsup#1,#2,{{\sl Prog.\ Theor.\ Phys.\/ Suppl.\/ }{\bf #1} (19#2)\ }
\def\rmp#1,#2,{{\sl Rev.\ Mod.\ Phys.\/ }{\bf #1} (19#2)\ }
\def\yadfiz#1,#2,#3[#4,#5]{{\sl Yad.\ Fiz.\/ }{\bf #1} (19#2) #3%
\ [{\sl Sov.\ J.\ Nucl.\ Phys.\/ }{\bf #4} (19#2) #5]}
\def\zh#1,#2,#3[#4,#5]{{\sl Zh.\ Exp.\ Theor.\ Fiz.\/ }{\bf #1} (19#2) #3%
\ [{\sl Sov.\ Phys.\ JETP\/ }{\bf #4} (19#2) #5]}

\def\beq{\begin{equation}}
\def\eeq{\end{equation}}
\def\beqar{\begin{eqnarray}}
\def\eeqar{\end{eqnarray}}

\def\nfrac#1#2{{\displaystyle{\vphantom1\smash{\lower.5ex\hbox{\small$#1$}}%
	\over\vphantom1\smash{\raise.25ex\hbox{\small$#2$}}}}}

\def\p#1{\mskip#1mu}

\def\stop{\p6.}
\def\comma{\p6,}

\def\f{\frac}

\def\l:{\mathopen{:}\,}
\def\r:{\,\mathclose{:}}

\def\[{\left[}          \def\]{\right]}
\def\({\left(}          \def\){\right)}
\def\<{\left<}          \def\>{\right>}

\begin{document}
\begin{titlepage}

\begin{center}
Feb 9, 1995\hfill       TAUP--2234--95 \\
\hfill                  hep-th/9502058

\vskip 1.5 cm
{\large \bf On Topological $2D$  String and Intersection Theory\\}
\vskip 1 cm
{Yaron Oz\footnote{Work
supported in part by the US-Israel Binational Science Foundation,
and GIF - the German-Israeli Foundation for Scientific Research.
}
}
\vskip 0.2cm

{\sl
yarono@ccsg.tau.ac.il \\
School of Physics and Astronomy\\Raymond and Beverly Sackler Faculty
of Exact Sciences\\Tel-Aviv University\\Ramat Aviv, Tel-Aviv 69978, Israel.
}

\end{center}

\vskip 0.5 cm
\begin{abstract}
The topological description of $2D$ string theory at the self-dual
radius is studied in the algebro-geometrical formulation of the
$A_{k+1}$ topological models at $k=-3$.
Genus zero correlators of tachyons and their gravitational
descendants are computed as intersection numbers on moduli space
and compared to $2D$ string results.
The interpretation of negative momentum tachyons as gravitational
descendants of the cosmological constant, as well as modifications of this,
is shown to imply a disagreement between $2D$ string correlators and
the associated intersection numbers.

\end{abstract}

\end{titlepage}

$2d$ string theory at the self-dual radius is described by a topological
field theory with an infinite number of primary fields.
A topological Landau-Ginzburg formulation of the theory
has been constructed in \cite{HOP,GM}, and a topological interpretation
of the $W_{1+\infty}$ recursion relations for the tachyon
correlators has been given
in \cite{LOS,GIM}.

However, the topological structure of the theory is still not well
understood. In particular, the identification of the BRST cohomology
states as gravitational primaries and descendants, and
the relation between the correlators and intersection numbers on the
moduli space of Riemann surfaces has not been clarified yet.

It has been suggested in \cite{HOP,GM} that while
non-negative momentum tachyons
correspond to primary fields, the negative momentum tachyons correspond
to
gravitational descendants of $T_0$, the zero momentum tachyon.
$2d$ string theory, however, is parity invariant under the reflection of
momenta $k \leftrightarrow -k$, thus positive and negative momentum
tachyons appear symmetrically. This might imply that
all the tachyons should be considered as primaries, which is not
in contradiction with the formulations of \cite{HOP,LOS}.
Such an interpretation is in agreement with intersection calculations
of the four-point tachyon correlator \cite{W1,MV} as well as the
$1 \rightarrow n$ tachyon amplitude and five-point function at
various kinematical regions \cite{DMP}.

There is a third possibility, namely that the negative
momentum tachyons can be considered
both as primaries and as descendants. This, if correct, will lead to
beautiful relations between intersection numbers on the moduli space of
Riemann surfaces, and should be intimately related to the way
gravitational descendants are built from matter \cite{LOSEV}.

Since the generating functional for tachyon correlators is a tau function
of the Toda lattice hierarchy \cite{DMP2},
one expects a generalization of the
relation that exists between intersection numbers and the
KdV hierarchy as implied
by topological gravity \cite{W2,KONS}.

In this letter we make a step in the study of the
intersection theory associated with the topological $2d$ string.
In particular, we study the possibility of relating negative momentum
tachyons
to gravitational descendants from the viewpoint of
intersection theory on the moduli space of Riemann surfaces.
We compute tachyon correlators as intersection numbers
and show that the identification of negative momentum tachyons as
gravitational descendants, as well as modifications of this,
is in disagreement with $2d$ string results.

The algebro-geometrical set up is given by the analytical continuation
of that, associated with the $A_{k+1}$ theory \cite{W1,W3}, to $k=-3$.
Define the "vector bundles"  $\nu$ and
$\omega$ over the compactified moduli space of genus $g$ Riemann
surfaces with $s$ punctures $\bar{{\cal M}}_{g,s}$, whose fibers  $V$ and $W$
respectively are \footnote{$\nu$ and $\omega$ are called the (higher) direct
image sheaves of ${\cal R}$ and are not necessarily locally free, i.e.
the dimension of the fiber may depend on the point on
$\bar{{\cal M}}_{g,s}$.}:
\beq
V = H^0(\Sigma_{g,s}, {\cal R}),~~~~~~~ W = H^1(\Sigma_{g,s}, {\cal R}) \comma
\label{fiber}
\eeq
where ${\cal R} \equiv K^2 \otimes_i O(z_i)^{1-k_i}$ is a
line bundle over $\Sigma_{g,s}$,
$K$ being the canonical bundle. The sections of ${\cal R}$
are two-forms that can have zeros at the points $z_i$
with order of at least $k_i-1$.

Then
\begin{equation}
\langle \prod_{i=1}^s \sigma_{n_i}(T_{k_i})\rangle_g
= \int_{\bar{{\cal M}}_{g,s}}
\prod_{i=1}^s (C_1({\cal L}_i))^{n_i}C_T(\nu \ominus \omega) \comma
\label{def}
\end{equation}
where $\sigma_n(T_k)$ is the n-th descendant of the momentum $k$ tachyon,
$C_T$ is the top Chern class and
$C_1({\cal L}_i)^n \in H^{2n}(\bar{{\cal M}}_{g,s})$ are
the Mumford-Morita-Miller stable classes. ${\cal L}_i$
is a line bundle on the moduli space with fiber at $\Sigma$
being the cotangent space to the point $x_i \in \Sigma$.
The subtraction of the vector bundles $\nu$ and $\omega$ should be understood
in
$K-$theory sense.

Using the Riemann-Roch theorem for a line bundle $L$
\begin{equation}
dim H^0(\Sigma_g,L) - dim H^1(\Sigma_g,L) = deg L - g + 1 \comma
\label{rr}
\end{equation}
and $deg(K) = 2g-2$, we get the selection rule for (\ref{def})
\begin{equation}
\sum_{i=1}^s (n_i - k_i) = 0 \stop
\label{sr}
\end{equation}

Let us consider
from the intersection theory viewpoint,
the definition of negative momentum
tachyons
$T_{-n}$ as the n-th gravitational descendants of $T_0$,
$T_{-n}= \sigma_n(T_0)$.
The first family of correlators we wish to evaluate is
$\langle \prod_{i=1}^s \sigma_{n_i}(T_{k_i})\rangle_0$ subject to
the constraint
$\sum_{i=1}^s n_i = s-3$ and satisfying (\ref{sr}).
Since $deg({\cal R}) = -1 < 0$, the
Kodaira vanishing theorem implies that
$dim V = 0$. Using the Riemann-Roch theorem (\ref{rr})
we get that $dim  W = 0$. Thus, in this case the correlator is given
by a simple integral
\begin{equation}
\langle \prod_{i=1}^s \sigma_{n_i}(T_{k_i})\rangle_0
= \int_{\bar{{\cal M}}_{0,s}}
\prod_{i=1}^s (C_1({\cal L}_i))^{n_i}
= \frac{(s-3)!}{\prod_{i=1}^s n_i!} \stop
\label{tc}
\end{equation}
This result is in contradiction with
$T_{-n}$ being defined as the n-th gravitational descendants of $T_0$,
since the correlators of $c=1$ tachyons satisfy
\begin{equation}
\langle \prod_{i=1}^s T_{k_i}\rangle_0  = 0 ~~~if~~~
\sum_{i=1}^s |k_i|\Theta(-k_i)= s-3  \comma
\label{relation}
\end{equation}
which is proved using the $T_0$ and $T_1$ Ward identities of the theory.

Note, however, that it is possible to overcome the disagreement between
(\ref{tc}) and (\ref{relation}) by postulating an infinite factor between
the tachyon $T_{-n}$ and the gravitational descendant $\sigma_n(T_0)$.

Consider next the genus zero three point function
\begin{equation}
\langle \prod_{i=1}^3 \sigma_{n_i}(T_{k_i})\rangle_0
= \int_{\bar{\cal M}_{0,3}}
\prod_{i=1}^3 (C_1({\cal L}_i))^{n_i}C_T(\nu \ominus \omega) \comma
\label{3pt}
\end{equation}
where $\sum_{i=1}^3 (n_i - k_i) = 0$.
$deg({\cal R}) = -\sum_{i=1}^3 k_i < 0$, thus
$dim V = 0$ and $dim  W = \sum_{i=1}^3 k_i$. The integral in (\ref{3pt})
is evaluated via
\begin{eqnarray}
\int_{\bar{\cal M}_{0,3}}
\prod_{i=1}^3(C_1({\cal L}_i))^{n_i}C_T(-\omega)  =
\int_{\bar{\cal M}_{0,3}}
\prod_{i=1}^3C_T({\cal L}_i^{\oplus n_i})C_T(-\omega) = \nonumber\\
=\int_{\bar{{\cal M}}_{0,3}}
C_T({\cal L}_{grav} \ominus \omega) =
\int_{\bar{{\cal M}}_{0,3}}
C_0({\cal L}_{grav} \ominus \omega) = 1 \comma
\label{id3pt}
\end{eqnarray}
where ${\cal L}_{grav} = \oplus_{i=1}^3{\cal L}_i^{\oplus n_i}$,
and we used a basic property of the top Chern class of a Whitney sum
of vector bundles
\begin{equation}
C_T(\nu \oplus \omega) = C_T(\nu)C_T(\omega) \stop
\end{equation}
Equation (\ref{id3pt}) is in agreement with the three point tachyon
correlator
\beq
\langle T_{k_1}T_{k_2}T_{k_3} \rangle = \delta_{k_1+k_2+k_3,0} \comma
\eeq
thus excluding the possibility of an infinite factor between
the negative momentum tachyon and the gravitational descendant.
Note that in contrast to (\ref{id3pt}),
in minimal topological models coupled to topological
gravity three point functions with gravitational descendants
vanish, which is one of the basic ingredients of
the topological recursion relations of \cite{W1}.

Consider now the genus zero four point function
\begin{equation}
\langle \prod_{i=1}^4 \sigma_{n_i}(T_{k_i})\rangle_0
= \int_{\bar{\cal M}_{0,4}}
\prod_{i=1}^4 (C_1({\cal L}_i))^{n_i}C_T(\nu \ominus \omega) \stop
\label{4pti}
\end{equation}
$deg({\cal R}) = -\sum_{i=1}^4 k_i < 0$, thus
$dim V = 0$ and $dim  W = \sum_{i=1}^4 k_i -1$. The integral in (\ref{4pti})
can be recast as
\begin{eqnarray}
\int_{\bar{\cal M}_{0,4}}
\prod_{i=1}^4(C_1({\cal L}_i))^{n_i}C_T(-\omega)  =
\int_{\bar{\cal M}_{0,4}}
\prod_{i=1}^4C_T({\cal L}_i^{\oplus n_i})C_T(-\omega) = \nonumber\\
=\int_{\bar{{\cal M}}_{0,4}}
C_T({\cal L}_{grav} \ominus \omega) =
\int_{\bar{{\cal M}}_{0,4}}
C_1({\cal L}_{grav} \ominus \omega) \comma
\label{id4pt}
\end{eqnarray}
where ${\cal L}_{grav} = \oplus_{i=1}^4{\cal L}_i^{\oplus n_i}$.

In order to evaluate (\ref{id4pt}) we will use the property that
the Chern character of a difference of vector bundles satisfies
\begin{equation}
ch(E \ominus F) = ch(E) - ch(F) \stop
\label{ch}
\end{equation}
The Chern Character has an expansion in terms of Chern classes
\begin{equation}
ch(E) = rank E + C_1(E) + \frac{1}{2}(C_1(E)^2 - 2C_2(E)) + ... \stop
\label{chern}
\end{equation}

Using (\ref{ch}) and (\ref{chern}) we have
\begin{equation}
ch({\cal L}_{grav} \ominus \omega) =
\sum_{i=1}^4 n_i(1+ C_1({\cal L}_i)) - (\sum_{i=1}^4
k_i-1 + C_1(\omega)+...) \stop
\label{id1}
\end{equation}
On the other hand, since ${\cal L}_{grav}\ominus \omega$
is a line bundle we get
\begin{equation}
ch({\cal L}_{grav} \ominus \omega) =
1+ C_1({\cal L}_{grav} \ominus \omega) \stop
\label{id2}
\end{equation}
Equations (\ref{4pti}),(\ref{id4pt}),(\ref{id1}) and (\ref{id2}) yield
\begin{eqnarray}
\langle \prod_{i=1}^4 \sigma_{n_i}(T_{k_i})\rangle_0
=\int_{\bar{{\cal M}}_{0,4}}
C_1({\cal L}_{grav} \ominus \omega) = \nonumber\\
= \sum_{i=1}^4n_i\int_{\bar{{\cal M}}_{0,4}}C_1({\cal L}_i) -
\int_{\bar{\cal M}_{0,4}}
C_1(\omega) =  \sum_{i=1}^4 n_i - \int_{\bar{{\cal M}}_{0,4}}C_1(\omega)
\stop
\label{ot}
\end{eqnarray}
Thus, in order to evaluate the genus zero four-point function of
tachyon's gravitational descendants we have to compute the first
Chern class of the vector bundle $\omega$.

Recall now that the four point tachyon correlator is \cite{kdf}
\beq
\langle T_kT_{k_1}T_{k_2}T_{k_3} \rangle =
(k-1) - \sum_{i=1}^3(k+k_i)\Theta(-k-k_i) \comma
\label{4pt}
\eeq
with $k > 0$.
As seen in (\ref{4pt}) the  $2\rightarrow 2$ amplitude
$\langle T_{-k_1}T_{-k_2}T_{k_3}T_{k_4}\rangle_0$ is non-trivial
and depends on various kinematical regions.
We will now compute this amplitude from intersection theory viewpoint
with negative momentum
tachyons considered as gravitational descendants of $T_0$.

Using (\ref{ot}) we have
\begin{equation}
\langle \sigma_{k_1}(T_0)\sigma_{k_2}(T_0)
T_{k_3}T_{k_4}\rangle_0=
k_1 +k_2 - \int_{\bar{{\cal M}}_{0,4}}C_1(\omega) \stop
\label{tt}
\end{equation}

In order to compute the first Chern class of the vector bundle $\omega$,
recall that
\begin{equation}
C_1(\omega) \equiv C_1(det ~\omega) \comma
\end{equation}
with $det ~\omega$ being the determinant line bundle associated with
$\omega$.
The rank of $\omega$ is $r = k_3+k_4-1$ and its fiber is the vector space
\beqar
H^1(\Sigma_{0,4}, K^2\otimes_{i=1}^2 O(z_i)^1
\otimes_{i=3}^4 O(z_i)^{1-k_i}) = \nonumber\\
H^0(\Sigma_{0,4}, K^{-1}
\otimes_{i=1}^2 O(z_i)^{-1}\otimes_{i=3}^4 O(z_i)^{k_i-1})^* \comma
\eeqar
where we used Serre duality
\begin{equation}
H^1(\Sigma_g,L) \cong H^0(\Sigma_g, K\otimes L^{-1})^*,
\end{equation}
and $*$ denotes the dual bundle.

A basis of $SL(2,C)$ invariant sections of the dual bundle to $\omega$ is,
for instance,
\begin{eqnarray}
s_1 & = & dz^{-1}\frac{(z-z_1)(z-z_2)}{z_{12}} \nonumber\\
s_i & = & dz^{-1}\frac{(z-z_1)^i(z-z_2)z_{13}^{i-1}}{(z-z_3)^{i-1}
z_{12}^i} ~~~~i=2...k_3 \nonumber\\
s_{k_3+i-1} \equiv s'_i & = &
dz^{-1}\frac{(z-z_1)^i(z-z_2)z_{14}^{i-1}}{(z-z_4)^{i-1}
z_{12}^i} ~~~~i=2...k_4 \comma
\label{section}
\end{eqnarray}
where $z_{ij} \equiv z_i-z_j$.
The first Chern class of $\omega$ is evaluated by counting with
multiplicities the zeros minus the poles of
a section $det(S)$ of $det~ \omega$. $S$ is the $r \times r$
matrix : $S_{ij} \equiv s_i(x_j)$.

The zeros and poles of $det(S)$ are located on the boundary
of $\bar{\cal M}_{0,4}$.
We denote by $\Sigma_{ij}$ the degenerate sphere corresponding
to the merge of the punctures $i$ and $j$, which in the Deligne-Mumford
compactification is represented by a splitting of a sphere.

The sections $s_i$ have a pole of order $i$ on $\Sigma_{12}$,
and a zero of order $i-1$ on $\Sigma_{13}$, while
$s'_i$ have a pole of order $i$ on $\Sigma_{12}$
and a zero of order $i-1$ on $\Sigma_{14}$, thus
contributing $k_3 + k_4 -1$ to the first Chern class
of $\omega$, where we used the fact that $C_1(\omega) =
-C_1(\omega^*)$.

We also have to calculate the zeros of the determinant due to linear
dependencies among the sections on $\Sigma_{ij}$.
Define
\beq
u(z) = \frac{z-z_1}{z-z_3}, ~~~~~v(z) = \frac{z-z_1}{z-z_4} \comma
\label{func}
\eeq
and consider the matrix $T_{ij}$

\beqar
T_{ij} & \equiv & u^i(x_j)       ~~~~~~~~~~~i=0...k_3-1,~~
{}~~~~~~~~j=1...k_3+k_4-1
\nonumber\\
& \equiv & v^{i-k_3+1}(x_j)   ~~~~i=k_3...k_3+k_4-2,~~ j=1...k_3+k_4-1
\stop
\label{T}
\eeqar
The matrix $T_{ij}$ is basically $S_{ij}$, but we factored out
overall non relevant terms.
It encodes the linear dependencies among the sections,
and we have to calculate the zeros of its determinant on the degenerate
surfaces $\Sigma_{ij}$.
Consider, for instance, the degenerate surface $\Sigma_{13}$.
{}From (\ref{func}) it follows that
\beq
u = \frac{\alpha}{1+ \f{\varepsilon}{v}} \comma
\label{v}
\eeq
where $\alpha = \frac{z_{14}}{z_{34}}$ and $\varepsilon =
\f{z_{13}}{z_{34}}$. We have to consider the order of zero of $T_{ij}$
as $\varepsilon$ goes to zero.
It is convenient to change a basis from $u^i, i=0...k_3-1$
to $(\f{\varepsilon}{v})^i$. Thus, the order of zero of $det(T)$
is $\sum_{i=1}^{k_3-1} = \f{1}{2}k_3(k_3-1)$.
Similarly we get the contributions from the other degenerate surfaces.
We find
\beq
det(T) \sim z_{13}^{\f{1}{2}k_3(k_3-1)}z_{14}^{\f{1}{2}k_4(k_4-1)}
z_{34}^{(k_3-1)(k_4-1)} \stop
\label{detT}
\eeq

Altogether we have
\begin{equation}
\int_{\bar{{\cal M}}_{0,4}}C_1(\omega) =
-\f{1}{2}(k_3+k_4)(k_3+k_4-5) - 2
\stop
\label{c1}
\end{equation}
Using (\ref{tt}) and (\ref{c1}) we get
\begin{equation}
\langle \sigma_{k_1}(T_0)\sigma_{k_2}(T_0)
T_{k_3}T_{k_4}\rangle_0=
k_1 +k_2 + 2+\f{1}{2}(k_3+k_4)(k_3+k_4-5)
\comma
\label{4ptans}
\end{equation}
where $\sum_{i=1}^4 k_i=0$.
For the special case
$\langle \sigma_1(T_0)\sigma_1(T_0)
T_1T_1\rangle_0$ only the section $s_1$ in (\ref{section})
is relevant, and the value of the
correlator is one, in agreement with (\ref{4ptans}).

The result (\ref{4ptans}) is clearly different from (\ref{4pt}),
in particular is does not depend on kinematical regions and it
is not parity invariant.

Note, in contrast, that if the negative momentum tachyons are considered as
primaries, then $dim V =1, dim W=0$, and the intersection calculation
\begin{equation}
\langle \prod_{i=1}^4T_{k_i}\rangle_0=
\int_{\bar{{\cal M}}_{0,4}}C_1(\nu) \comma
\label{tti}
\end{equation}
agrees with (\ref{4pt}) \cite{W1}.

In \cite{GIM} a definition of a negative momentum tachyon
as a sum of gravitational descendants in $1\rightarrow n$
amplitudes has been proposed
\begin{equation}
T_{-n} = \sum_{i=0}^n \prod_{j=1}^i (j-n) \sigma_i(T_{-n+i})\stop
\end{equation}
This definition together with (\ref{def})
is clearly compatible with the $1\rightarrow 3$
tachyons correlator.
However, for the $2\rightarrow 2$ amplitude we get
$dim V = dim W =0$, thus
\begin{eqnarray}
\langle T_{-k_1}T_{-k_2}T_{k_3}T_{k_4}\rangle_0 =
(1-k_1)\int_{\bar{{\cal M}}_{0,4}}
C_1({\cal L}_1) +
\nonumber\\
+(1-k_2)\int_{\bar{{\cal M}}_{0,4}}
C_1({\cal L}_2) = 2 - k_1 -k_2 ,
\end{eqnarray}
in disagreement with (\ref{4pt}).

The intersection theory computations presented above imply that
the identification of
negative momentum tachyons as gravitational descendants
in the topological formulation of the $2d$ string
leads to a disagreement between
tachyon correlators and the associated intersection numbers.
The identification of all the tachyons as gravitational primaries
is consistent with all the intersection theory computations which have been
done so far.
These lead us to suggest that indeed the proper set up is to consider
all the tachyons as gravitational primaries, and interpret accordingly
the topological numbers associated with their correlators.

It should be clear, however, that our computations do not exclude the
possibility that an insertion of a tachyon
operator in a correlator might be equivalent to
an insertion of some combination of gravitational
descendants. This would simply mean that there are topological
Ward identities underlying the theory.

The gravitational descendants of the tachyons, which some of their
correlators have been calculated above, are likely to correspond
to discrete states of $2d$ string theory.
However, as discussed in \cite{ho}, it is not clear what is the exact
relation between gravitational
descendants and discrete states since correlators of the latter
have not been
calculated successfully in the Liouville framework.
Such a relation, when revealed, will evidently shed light on the
relation between the physical $2d$ string theory and its topological
formulation. Also, an underlying
integrable structure generalizing the Toda lattice
hierarchy is expected.

It is straightforward to apply the same manipulations which
we used in order to derive  (\ref{id3pt}) and (\ref{ot}), in order to get
formulas for higher point functions of tachyons and
their gravitational descendants.
These formulas involve higher Chern classes of the sheaf
$\omega$. The latter is not locally free and we expect jumps in the
dimension of the fiber at the boundary of the moduli space
\footnote{Such phenomena appeared in tachyons five-point
computations of \cite{DMP}.}, thus
complicating the calculation of the intersection numbers.
A proper application of the Grothendieck-Riemann-Roch theorem is likely
to be helpful for these computations.

Finally, we hope that the ideas used in the
calculations that we made will be useful for the complete study of
topological $2d$ string theory.

{\bf Acknowledgments} We would like to thank J. Bernstein and R. Plesser
for helpful discussions.


\newpage
\begin{thebibliography}{99}

\small
\parskip=0pt plus 2pt

\bibitem{HOP} A.~Hanany, Y.~Oz and R.~Plesser,
``Topological Landau-Ginzburg Formulation and Integrable Structure
of 2d String Theory,''hep-th/9401030, \np425,94,150.
\bibitem{GM} D.~Ghoshal, S.~Mukhi, ``Topological Landau-Ginsburg
Model of Two-Dimensional String Theory,''
hep-th/9312189, \np425,94,173.
\bibitem{LOS} Y.~Lavi, Y.~Oz and J.~Sonnenschein, ``$(1,q=-1)$ Model as
a Topological Description of $2d$ String Theory,''
hep-th/9406056, \np431,94,223.
\bibitem{GIM}D.~Ghoshal, C.~Imbimbo and S.~Mukhi, ``Topological 2D String
 Theory: Higher-genus Amplitudes and W-infinity Identities,'' MRI-PHY/13/94,
 CERN-TH-7458/94, TIFR/TH/39-94,
hep-th/9410034.
\bibitem{W1} E.~Witten,
``The $N$ Matrix Model and Gauged WZW Models,'' \np371,92,191.
\bibitem{MV} S.~Mukhi and C.~Vafa, ``Two Dimensional Black
Hole As a Topological Coset Model of $c=1$ String Theory,''
hep-th/9301083, \np407,93,667.
\bibitem{DMP} R.~Dijkgraaf, G.~Moore and R.~Plesser, unpublished.
\bibitem{LOSEV} A.~Lossev, ``Descendants Constructed From Matter Fields
in Topological Landau-Ginzburg Theories Coupled to Topological Gravity,''
ITEP Preprint PRINT-92-0563 (Jan 1993), hepth/9211089.
\bibitem{DMP2} R.~Dijkgraaf, G.~Moore and R.~Plesser, ``The partition
function of 2D string theory,'' hep-th/9208031, \np394,93,356.
\bibitem{W2} E.~Witten,
``Two Dimensional Gravity and Intersection Theory on
Moduli Space,'' In Cambridge 1990, Proceedings, Surveys in Differential
Geometry 243-310.
\bibitem{KONS} M.~Kontsevich, ``Intersection Theory on The Moduli Space
of Curves and The Matrix Airy Function,'' \cmp147,92,1.
\bibitem{W3} E.~Witten,
"Algebraic Geometry Associated With Matrix Models of Two
Dimensional Gravity", IASSNS-HEP-91/74.
\bibitem{kdf} P.~di~Francesco, D.~Kutasov,
"World Sheet and Space-Time Physics in Two Dimensional (Super)
String Theory", hep-th/9109005, \np375,92,119.
\bibitem{ho} A.~Hanany and Y.~Oz,
``$c=1$ Discrete States Correlators via $W_{1+\infty}$ Constraints,''
hep-th/9410157, To appear in {\em Phys.Lett.B}.

\end{thebibliography}
\end{document}